\documentclass[prd,twocolumn,showpacs]{revtex4}
\usepackage{bm}
\usepackage{amssymb}
\begin{document}
\title{Instability of scalar charges in 1+1D and 2+1D}
\author{Lior M.\ Burko}
\affiliation{Department of Physics, University of Utah, Salt Lake City,
Utah 84112}
\date{\today}
\begin{abstract}
We study the phenomenon of mass loss by a scalar charge --- a point
particle which acts as a source for a (non-interacting) scalar 
field --- in ($1+1$)-dimensional and ($2+1$)-dimensional flat
spacetime. We find that such particles are unstable against self
interaction: they lose their mass through the emission of monopole
radiation. This is in sharp contrast with scalar charges in
($3+1$)-dimensional flat spacetime, where no such phenomenon occurs. 
\end{abstract}
\maketitle

\section{Introduction and summary}

The absence of observations of scalar charges --- pointlike particles
which act as sources of non-interacting scalar
fields --- is an interesting fact, in particular because of the simplicity
of the scalar field theory. This raises the question of whether 
the model of scalar charges is self consistent. Specifically, one can ask
whether scalar charges are stable against self interaction. As we shall
show in this paper, scalar charges in ($1+1$)-dimensions and in 
($2+1$)-dimensions cannot have a constant rest 
mass already in flat spacetime. Instead, they lose their mass through the
emission of monopole 
waves. 

The variable rest mass is an intersting phenomenon, specific to
scalar charges, which surprisingly has attracted only little attention,
despite its being noted long ago \cite{nordstrom}. 
Recently, the mass loss by a scalar charge has been studied in the
spacetime of an expanding universe \cite{burko-harte-poisson}. It is shown
in Ref.\ \cite{burko-harte-poisson} that scalar charges must have variable
mass in a class of cosmological spacetimes, which includes the de Sitter
spacetime and a spatially-flat matter dominated cosmology. It is also
shown that the properties of such particles are invariably coupled to the
cosmological parameters. In this paper we shall study a similar effect in
flat spacetime in ($1+1$)-dimensions and in ($2+1$)-dimensions. 

Interestingly, similar phenomena never happen with electric 
charges. Specifically, the force on an electric charge $e$ is given by the
Lorentz force law, $f_{\alpha}=eF_{\alpha\beta}u^{\beta}$, where
$F_{\alpha\beta}$ is the Maxwell field-strength tensor, and $u^{\beta}$ is
the particle's four velocity. In the rest frame of the particle this
reduces to $f_{\alpha}=eF_{\alpha t}$, such that the $t$ component has to
vanish because of the skew symmetry of Maxwell's tensor. 

Because of the lack of observations, the scalar field theory is not
unique. In particular, there is a number of possible ways to 
couple sources to scalar fields. Some of the action principles ---
notably action principles which lead to nonlinear couplings --- do not
share the phenomenon of variable rest mass \cite{mtw,burko-harte-poisson}. 
We use here the simplest action principle, which is also the most popular
one. This action principle leads to the simplest, linear wave equation. It
also yields a force law in which the force is proportional to the field's
gradient. This implies that the force is not necessarily orthogonal to the
particle's world line. Our results are restricted to this action
principle. Although other action principles may be interesting to
consider, it is interesting to study the implications of a 
given theory, especially because it is an unusual one. Further, similar
phenomena in realistic spacetimes may provide an explanation to the lack
of observations of such particles \cite{burko-harte-poisson}. 

It is interesting to note that in ($3+1$)-dimensions this phenomenon
occurs only in certain classes of cosmological spacetimes (including 
de Sitter spacetime and spatially-flat matter dominated cosmology
\cite{burko-harte-poisson}), but it does not occur in wide classes of
other spacetimes, including spacetimes of stationary black holes 
\cite{burko-liu}. It is easy and instructive to see why nothing
interesting happens in ($3+1$)-dimensions Minkowski spacetime. The
retarded Green's function in ($3+1$)-dimensions for a source at
the origin of the coordinates is given by 
$G(t,r;t',0)=\delta[(t-t')-r]/r$. Because the Green's function has support
only on the light cone, there is no ``tail'', or wake, which has support
inside the light cone. Consequently, the field $\Phi$ of a static source
is strictly static, such that the temporal component of its derivative
vanishes. The situation in ($2+1$)-dimensions and ($1+1$)-dimensions 
is markedly different already in flat spacetime. The Green's function has
support inside
the light cone, and this leads to non-trivial self interaction.

This paper is organized as follows. In Section \ref{eom} we write the
action principle and the equations of motion. 
In Section \ref{sec2} we study the
self interaction in ($1+1$)-dimensions, and in Section \ref{sec3} we
study it in ($2+1$)-dimensions. We find that scalar particles in either
($1+1$)-dimensions or ($2+1$)-dimensions must lose their rest mass. The
lost energy is radiated to infinity, in addition to some of the energy
which is stored in the field at finite distances, such that globally
energy is conserved.

\section{action principle and equations of motion} \label{eom}

We take the action principle in $D$ dimensions
to be \cite{burko-harte-poisson}
\begin{eqnarray}
S&=&\int\left\{-\frac{1}{8\pi}g^{\alpha\beta}\Phi_{,\alpha}\Phi_{,\beta}-\int
(E_0-q\Phi)\right. \nonumber \\
&\times& \left.
\sqrt{-g_{\alpha\beta}
\frac{\,dz^{\alpha}}{\,d\lambda}\frac{\,dz^{\beta}}{\,d\lambda}}\frac{\delta^D
[x,z(\lambda)]}{\sqrt{-g}}\,d\lambda\right\}\sqrt{-g}\,d^Dx
\end{eqnarray}
where $E_0$ is the particle's bare mass, $q$ is the scalar charge which is
the source for the scalar field $\Phi$, $g_{\alpha\beta}$ is the fixed
metric of the background whose determinant is $g$, and $\lambda$ is an
arbitrary parametrization of the particle's world line
$z^{\alpha}(\lambda)$. 
Variation of the action with respect to $\Phi$ yields the wave equation
\begin{equation}
\nabla_{\mu}\nabla^{\mu}\Phi(x^{\alpha})=-4\pi\rho(x^{\alpha})\, ,
\label{wave-eq}
\end{equation}
where the scalar charge density is given by
\begin{equation}
\rho(x^{\alpha})=\int q(\tau)\frac{\delta^D[x,z(\lambda)]}
{\sqrt{-g}}\,d\tau\, ,
\end{equation}
and $\tau$ is the particle's proper time. Variation of the action with
respect to the world line yields the force law
$f_{\alpha}=q\Phi_{,\alpha}$. Notice that the force $f_{\alpha}$ is not
necessarily orthogonal to the world line. This implies that the mass of
the particle does not have to be conserved. 

Identical force law and wave equation can be obtained also
from alternative action principles. For example, some authors use the
action \cite{quinn}
\begin{eqnarray}
S' &=& \int \biggl\{ -\frac{1}{8\pi} g^{\alpha\beta} \Phi_{,\alpha}
\Phi_{,\beta}   
+ \int \biggl[ \frac{1}{2} m(\tau)\, g_{\alpha\beta} u^\alpha u^\beta
\nonumber \\
& & \mbox{}   
+ q \Phi \biggr]\, \frac{\delta^D[x,z(\tau)]}{\sqrt{-g}}\, d\tau
\biggr\} \sqrt{-g}\, d^D x \, .
\end{eqnarray}
Note, however, that this action is not invariant under reparametrization
of the world line. 

We note that it is easy to construct action principles which lead to
equations of motion where the force is always orthogonal to the world
line. For example, consider the action principle \cite{mtw} 
\begin{eqnarray}
S''&=&\int\left\{-\frac{1}{8\pi}g^{\alpha\beta}\Phi_{,\alpha}\Phi_{,\beta}-\int
E_0\exp(-q\Phi/E_0)\right. \nonumber \\
&\times& \left.
\sqrt{-g_{\alpha\beta}
\frac{\,dz^{\alpha}}{\,d\lambda}\frac{\,dz^{\beta}}{\,d\lambda}}\frac{\delta^D
[x,z(\lambda)]}{\sqrt{-g}}\,d\lambda\right\}\sqrt{-g}\,d^Dx\, .
\end{eqnarray}
Variation of $S''$ with respect to the world line yields
\begin{equation}
E_0 \frac{Du^{\mu}}{\,d\tau}=q(g^{\mu\nu}+u^{\mu}u^{\nu})\Phi_{,\nu}\, ,
\label{orth-eom}
\end{equation}
which is explicitly orthogonal to the world line. Variation of $S''$ with
respect to the field $\Phi$ yields the wave equation
\begin{equation}
\nabla_{\mu}\nabla^{\mu}\Phi=-4\pi\rho\exp(-q\Phi/E_0)\, .
\label{mod-we}
\end{equation}
The wave equation (\ref{mod-we}) includes a source term which is coupled
nonlinearly to the field $\Phi$. We note, that several authors have used
the projected force law (\ref{orth-eom}) in tandem with the wave equation 
(\ref{wave-eq}) [instead of (\ref{mod-we})]. This choice is obviously
inconsistent \cite{galtsov-ori-burko}.

\section{($1+1$)-D flat spacetime}\label{sec2}

The metric is given simply by 
\begin{equation}
\,ds^2=-\,dt^2+\,dx^2 \; .
\end{equation}
The scalar field equation with this metric is given by Eq.\
(\ref{wave-eq}), 
where $\nabla_{\mu}\nabla^{\mu}=-\partial_{t}^2+\partial_{x}^2$, and where
the charge density 
$\rho(t,x)=q(t)\delta(x)\Theta(t-t_0)$. 

Normally, one may expect the charge $q$ to be a conserved quantity. For
scalar
charges, however, no such conservation law exists, and thus we generally
allow $q$ to vary with the time $t$. 
Also, we assume here that the charge density vanished for
$t<t_0$. (Later, we can take $t_0\to -\infty$.) For simplicity we assume
that the charge is static, because the phenomenon in interest occurs
already there, and we place the charge at the center of the coordinates
without loss of generality. 

One could argue the following: Spacetime is static, the source for the
field is static, and therefore we could expect the field itself to satisfy
the same symmetry, i.e., be static. This argument fails for two
reasons. First, the
charge was born at $t=t_0$, hence it is not strictly static. However,
even if $t_0\to -\infty$, the static solution is untenable. To illustrate
this point, let us assume for now that $q(t)=q_0={\rm const}$. 
It is simple
to find the static solution, which is given by $\Phi_{\rm static}(x)=-2\pi
q_0 
|x|$. 
The problem with this solution is that, in a certain sense, it requires
infinite energy. To see that let us consider scalars built from the
stress-energy tensor. Two possible scalars are $T\equiv
(T_{\alpha\beta}T^{\alpha\beta})^{1/2}$ and ${\cal T}\equiv
T_{\alpha}^{\alpha}$. In any dimensions $D$, ${\cal T}=\frac{1}{8\pi}
(2-D)\Phi_{,\alpha}\Phi^{,\alpha}$, and 
$T=\frac{1}{8\pi}D^{1/2}\Phi_{,\alpha}\Phi^{,\alpha}$. As in
($1+1$)-dimensions ${\cal T}$ is always zero, we shall consider here
$T$. With the simple static solution, $\Phi_{,\alpha}\Phi^{,\alpha}=4\pi
q_0^2$, which never decays. Namely, even if we go to infinity, the
stress-energy never drops off, such that its integral over the entire
space diverges. Note, that unlike the infinite potential energy of a
classical electron, this divergence comes from the behavior of the field
at infinity, and not from the extrapolation of the field to the
coincidence limit with its source. 
The reason for this ill behavior is that with the static solution
we did not
require appropriate boundary conditions at infinity. 

In order to find the
correct physical solution, let us require that the boundary conditions
are that there is no incoming radiation at infinity. Then the
retarded Green's function is given by \cite{morse-feshbach}
\begin{equation}
G(t,x;t',x')=2\pi\Theta(t-t'-|x-x'|) \, .
\end{equation} 
An interesting property of this Green's function is that it has support
inside the light cone. (Note, that one can always add a constant to the
Green's function. In particular, one can make $G$ vanish inside the light
cone. Then, however, $G$ becomes non-zero outside the light cone, in such
a way that the solution for the field is unchanged.) 
The field is obtained by convolving the source with the Green's
function. Specifically, 
\begin{eqnarray}
\Phi(t,x)&=&\int_{-\infty}^{\infty}\int_{-\infty}^{\infty}\,dt'\,dx'
\rho(t',x')G(t,x;t',x')\nonumber \\
&=& 2\pi[Q(t-|x|)-Q(t_0)]
\end{eqnarray}
where $Q(z)\equiv\int^{z}q(x)\,dx$. Note, that this solution
is time dependent. Recall, that the assumption of no ingoing radiation at
infinity is implicit in the Green's function. This assumption breaks the
time invariance of the problem, even though we do not stipulate outgoing
radiation at infinity. We remark that this solution has no tail: 
the field propagates strictly along the characteristics. It is the time
dependence which leads to the effect in consideration.
Next, we compute $\Phi_{,\alpha}\Phi^{,\alpha}$, and find that it vanishes
identically. Consequently, this solution does not share the problems of
the static solution.  

We next study the self interaction of the source for the field, the charge
$q$. The self force acting on this charge is given simply by
$f_{\alpha}(t)=q(t)\partial_{\alpha}\Phi(t,x)|_{x=0}$. Substituting the
solution for the field we find that 
\begin{equation}
f_{\alpha}=2\pi q^2(t)\delta^{t}_{\alpha}\, .
\end{equation}
This self force has only a temporal component. From symmetry, it is clear
why the spatial component has to be zero. However, the temporal component
is non-zero in the {\it rest frame} of the charge. This implies that the
mass of the particle cannot remain constant: it must vary with time. As 
$f_t\equiv\,dp_t/\,dt=-\,dE/\,dt$, we find that $\,dE/\,dt=-2\pi q^2(t)$,
which
is a negative-definite quantity. Here, $p_{\mu}$ is the particle's
four-momentum, and $E$ is its energy. Consequently, the mass of the particle
necessarily decreases with time. 

In the
simplest case, where the energy of the particle has no scalar-charge
origin, the charge $q=q_0$ does not vary with time. The particle's rest
energy is then given by $E(t)=E_0-2\pi q_0^2t$, which decreases linearly
with time. At the time $\tau=E_0/(2\pi q_0^2)$ the energy of the particle
vanishes, but $\dot{E}$ is not zero then. Strictly speaking, the particle
could continue to lose mass, and at times $t>\tau$ the mass of the
particle would become negative, and continue to grow ever more negative.  
In order to avoid such runaway problems, we recall that our purely
classical treatment cannot predict what happens when a particle loses all
its mass. A way to resolve this difficulty is the
following. Recall that there is no charge conservation for scalar
charges. That is, the charge does not have to be constant. One could
consider models in which $q$ varies with time, such that $E,\dot{E}$ and
$q$ vanish simultaneously. In such models the particle just ceases to
exist when both its charge and rest mass vanish. Alternative possibilities
are that the charge vanishes before the mass does. This is the case when
$E_0$ is very large. In that case we would be left with an uncharged
remnant. However, also the original possibility could have an interesting
implication. Because of the non-conservation of scalar charges, the charge
could abruptly disappear at the time that $E$ vanishes, such that no
negative-energy particles would be produced \cite{burko-harte-poisson}. 

Two
simple models for a
scalar-field origin of the particle's mass are the following. In both
models
we assign to the particle at time $t_0$ a charge $q_0$. 
First, in analogy with 
the classical electron models, we can introduce a new length scale, the
classical radius of a scalar particle, and postulate that the particle's
mass $E=\alpha q^2$. This implies that
$q(t)=q_0\exp[-(\pi/\alpha)(t-t_0)]$. Second, a simpler model, which does
not
require the introduction of a new length scale, is the one where the
particle's mass is proportional to its charge, i.e., $E=\beta |q|$. (As
motivation for this model consider the case of a charged black hole, for
which the maximal mass is equal to its charge.) This implies that 
$q(t)=q_0/[1+\frac{2\pi}{\beta}q_0(t-t_0)]$. In both models the particle's
charge and mass vanish only at infinitely late times. 

Although the local energy of the particle decreases with time, global
energy is still conserved. Naturally, we expect energy flux to
infinity. Indeed, we find that that flux is given by
\begin{eqnarray}
{\cal F^{\pm}}&=&-T_t^x \nonumber \\
&=& \pi q^2(t-|x|) (x/|x|)\, ,
\end{eqnarray}
where ${\cal F}^{\pm}$ is the flux to the positive or negative
directions, respectively. The total radiated flux
is given then by
${\cal F}={\cal F}^+-{\cal F}^-=2\pi q^2(t-|x|)$. It is interesting to
note that $\dot{E}+{\cal F}\ne 0$, except for the special case where the
charge $q$ is constant. This is the case because of a retardation
effect: The lost mass depends on the charge evaluated at a time $t$,
whereas the flux depends on the charge evaluated at the time
$t-|x|$. Unless the charge does not change with time, these two quantities
will not balance each other to guarantee global energy conservation. In
fact, we find that the flux to infinity is greater than the rate at which
the particle loses mass. The extra flux comes from the energy stored in
the field at finite distances, which decreases with time. 
To find the rate of change of the energy stored in the field, we compute 
\begin{equation}
\frac{d}{\,dt}\int_{-x}^{x}T^{tt}\,dx=-2\pi[q^2(t-x)-q^2(t)]
\end{equation}
where $x>0$. Now, we do indeed find that 
\begin{equation}
\frac{d}{\,dt}E+{\cal F}+\frac{d}{\,dt}\int_{-x}^{x}T^{tt}\,dx=0
\end{equation}
such that global energy is explicitly conserved. Note that this is the
case for any value of $x$, and in particular for $x\to\infty$, i.e., the
entire space. 

\section{($2+1$)-D flat spacetime}\label{sec3}

The metric is now
\begin{equation}
\,ds^2=-\,dt^2+\,dr^2+r^2\,d\theta^2\, 
\end{equation}
such that the field equation  is given by Eq.\ (\ref{wave-eq}), where the
wave operator is
$\nabla_{\mu}\nabla^{\mu}=-\partial_t^2+\partial_r^2+(1/r)\partial_r$, and
the charge density is given by 
$\rho=q(t)\delta(r)\Theta(t-t_0)$. For brevity we shall consider here only
the case of constant charge, i.e., $q=q_0={\rm const}$. [Generalization to
variable charge is analogous to the case of ($1+1$)-dimensions.] In 
analogy with ($1+1$)-dimensions, we
seek boundary conditions with no incoming radiation from infinity. The
Green's function then is given by \cite{morse-feshbach}  
\begin{equation}
G(t,r;t',r')=\frac{2\Theta[(t-t')-(r-r')]}{\sqrt{(t-t')^2-(r-r')^2}}\, .
\end{equation}
Convolving the charge density with this Green's function we find that 
\begin{equation}
\Phi(t,r)=2q_0\ln\left[\frac{r}{(t-t_0)-\sqrt{(t-t_0)^2-r^2}}\right]\, .
\end{equation}  
In ($2+1$)-dimensions the solution has support inside the light cone, and
it is time dependent. It is indeed a well known phenomenon, that in odd
spacetime dimensions the Huygens principle is violated. 
Notice that $\Phi$ vanishes on the light cone, and
that it is undefined outside the light cone. 
This solution satisfies $\Phi_{,\alpha}\Phi^{,\alpha}=4q_0/r^2$. The self
force acting on the particle is $f_t=2q_0^2/(t-t_0)$, such that the rate
at which mass is dissipated is given by ${\dot E}=-2q_0^2/(t-t_0)$. Notice
that any finite mass particle must had infinite mass at the time
$t_0$. Assuming that at the time $t_1>t_0$ the particle had mass $E_1$, we
find that $E(t)=E_1-2q_0^2\ln [(t-t_0)/(t_1-t_0)]$, such that at the
finite time $\tau=t_1\exp[E_1/(2q_0)]$ the particle loses all its mass.

Again, global energy is conserved: The flux through a circle of radius $r$  
is given by 
\begin{eqnarray}
{\cal F}&=&-\int_{0}^{2\pi}T_t^rr\,d\theta\nonumber \\
&=&2q_0^2\frac{t-t_0}{(t-t_0)^2-r^2}\, .
\end{eqnarray}
As in the case of ($1+1$)-dimensions, the flux to infinity is greater
than
the rate at
which
the particle loses mass. The rest of the flux comes from a decrease in the
energy stored by the field, at the rate of
\begin{equation}
\frac{d}{\,dt}\int_{0}^{2\pi}\,d\theta\int_{0}^{r}\,drT^{tt}r
=-2q_0^2\frac{r^2}{[(t-t_0)^2-r^2](t-t_0)}\, ,
\end{equation}
such that 
\begin{equation}
\dot{E}+{\cal F}+\frac{d}{\,dt}\int_{0}^{2\pi}\,d\theta\int_{0}^{r}\,dr
T^{tt}r=0\, .
\end{equation}
This energy conservation is valid for any radius $r$, and in particular
for $r\to\infty$. 

\section{Conclusions}

We have shown that scalar charges cannot maintain constant mass in
($1+1$)-dimensions and in ($2+1$)-dimensions, already in flat spacetime. 
In fact, the mass of the particle must always decrease with time. 
The particle loses its mass because of its self interaction, through
the mechanism of emission of monopole waves. In this sense scalar charges
are unstable against self interaction. In ($3+1$)-dimensions, however,
scalar charges are stable in flat spacetime, and also in certain
non-cosmological spacetimes, e.g., in black hole spacetimes. The situation
is different when cosmological spacetimes (e.g., de Sitter spacetime, or a
matter dominated cosmology) are considered \cite{burko-harte-poisson},
where a similar mass loss is found. 

We remark that we have only studied this effect classically. It is
interesting to see how scalar particles behave quantum mechanically or 
in dimensions greater than four.

\begin{acknowledgments}

I thank Abraham Harte, Poghos Kazarian, Karel Kucha\v{r}, Amos Ori, and
Richard Price for
discussions. Initial
work on this project was done  
at the California Institute of Technology, where it was
supported by  NSF grants AST-9731698 and PHY-9900776, and by NASA grant
NAG5-6840. This research was supported at the University of Utah by NSF
grant PHY-9734871.

\end{acknowledgments}

\end{document}